\documentclass[reprint,superscriptaddress,secnumarabic,amssymb,aps,prx,nobalancelastpage]{revtex4-2}

\newcommand{\backtopic}[1]{\par\addvspace{5pt}\noindent\textit{#1.}\enspace\ignorespaces}

\ifdefined\pdfinclusioncopyfonts
\fi

\usepackage{placeins}
\usepackage{capt-of}

\usepackage{graphicx}
\usepackage{amsmath,amssymb}
\usepackage{braket}
\usepackage[colorlinks,linkcolor=blue,citecolor=blue,urlcolor=blue,filecolor=black]{hyperref}

\usepackage[dvipsnames]{xcolor}
\usepackage{tikz}
\usepackage{enumerate}

\usetikzlibrary{arrows.meta, positioning, calc, fit, backgrounds, decorations.pathreplacing, patterns}

\definecolor{deepslate}{HTML}{355070}
\definecolor{warmrust}{HTML}{B7653A}
\definecolor{sage}{HTML}{81B29A}
\definecolor{wheat}{HTML}{F2CC8F}
\definecolor{ocean}{HTML}{3D7EA6}
\definecolor{mauve}{HTML}{6D597A}

\tikzset{
    pipelinebox/.style={
        draw=deepslate!50, fill=deepslate!5,
        rounded corners=3pt, minimum height=1.0cm, minimum width=2.0cm,
        font=\sffamily\small\bfseries, align=center,
        inner sep=5pt, line width=0.7pt,
    },
    convbox/.style={
        draw=warmrust!60, fill=warmrust!8,
        rounded corners=3pt, minimum height=1.0cm, minimum width=2.0cm,
        font=\sffamily\small\bfseries, align=center,
        inner sep=5pt, line width=0.7pt,
    },
    pipearrow/.style={
        -{Stealth[length=6pt, width=4.5pt]},
        line width=0.9pt, deepslate!60,
    },
    branchline/.style={
        line width=0.7pt, deepslate!35, densely dashed,
    },
    geolabel/.style={
        font=\sffamily\small\bfseries, align=center, text=deepslate!80,
    },
    geosublabel/.style={
        font=\sffamily\scriptsize, align=center, text=deepslate!50,
    },
    panellabel/.style={
        font=\sffamily\bfseries\normalsize,
    },
    paneltitle/.style={
        font=\sffamily\small, text=deepslate!70,
    },
}

\usepackage{setspace}
\usepackage[export]{adjustbox}
\usepackage{dsfont}
\usepackage[capitalize]{cleveref}
\crefname{appendix}{Appendix}{Appendices}
\Crefname{appendix}{Appendix}{Appendices}
\crefname{edfigure}{Extended Data Fig.}{Extended Data Figs.}
\Crefname{edfigure}{Extended Data Fig.}{Extended Data Figs.}
\crefname{edtable}{Extended Data Table}{Extended Data Tables}
\Crefname{edtable}{Extended Data Table}{Extended Data Tables}
\usepackage{bm}
\usepackage{mathtools}
\usepackage{url}
\usepackage{enumitem}
\usepackage{MnSymbol}
\usepackage{physics}
\usepackage{stmaryrd}
\usepackage{booktabs}
\usepackage{multirow}
\makeatletter\newcommand{\rulecolor}[1]{\noalign{\gdef\CT@arc@{\color{#1}}}}\makeatother

\begin{document}

\title{Scaling classical hardware for accurate real-time quantum error correction}

\author{Chenyang (Eric) Jin}
\affiliation{Department of Electrical Engineering and Computer Sciences, University of California, Berkeley, California 94720, USA}
\author{Susanne F. Yelin}
\affiliation{Department of Physics, Harvard University, Cambridge, Massachusetts 02138, USA}
\author{Andi Gu}
\email{andigu@g.harvard.edu}
\affiliation{Department of Physics, Harvard University, Cambridge, Massachusetts 02138, USA}

\begin{abstract}
Fault-tolerant quantum computation requires a classical decoder fast enough to process syndrome measurements at the rate at which they are generated, and to return its decisions with low latency.
This is most demanding for superconducting processors, whose fast syndrome cycles force a trade-off between decoding speed and accuracy.
Decoders fast enough to keep up sacrifice accuracy, which must be compensated for with larger codes and thus additional physical qubits.
Here we introduce \emph{Rapids}, a streaming neural decoder that is both fast and accurate.
We achieve this by co-designing the decoder with a field-programmable gate array (FPGA): the network is built from operations that map cheaply onto the device and trained in its integer arithmetic.
For the distance-19 surface code, Rapids decodes each round of syndromes in under a microsecond on a single FPGA, fast enough for superconducting processors, and returns its decisions with the low latency that fault-tolerant algorithms require.
Its logical error rate is 240 times lower than that of existing real-time hardware decoders.
The same design extends to the color code and the $\llbracket 144,12,12\rrbracket$ bivariate-bicycle code, on which our decoder is 16 times more accurate than the real-time Relay decoder.
We further show that the decoder scales: spread over several FPGAs or hardened into custom silicon, it becomes faster, more accurate, or able to decode larger codes.
\emph{Classical} hardware thus becomes a further resource, alongside qubits, that can be scaled to improve fault-tolerant quantum computers.
\end{abstract}
\maketitle

A key component of every fault-tolerant quantum computer is the decoder.
This decoder must meet stringent speed requirements, in both throughput (how many rounds of syndromes it processes per second) and latency (how long it takes to return a decision once the relevant syndromes have arrived).
Its throughput must keep up with the rate at which syndromes are measured on the quantum device: a decoder that falls even slightly behind accumulates a backlog that grows without bound, slowing the computation exponentially~\cite{terhal2015quantum,skoric2023parallel}.
The decoder must also have a low decision latency, because the runtime of many quantum algorithms is directly limited by this latency~\cite{fowler2013timeoptimalquantumcomputation,litinski2019game,gidney2021factor,beverland2022assessing,gidney2025factor}. 
Finally, it must be accurate. Its logical error rate determines the code distance a computation requires, and a less accurate decoder must be compensated for with larger codes and thus additional physical qubits.
Meeting all three at once is difficult, and hardest on superconducting processors, which complete a round of syndrome extraction approximately every microsecond~\cite{google2024willow}.
So far, decoders have kept pace with this rate only by sacrificing much of their accuracy~\cite{caune2024demonstrating,liyanage2024fpga,barber2025realtime,ziad2025local,muller2024relax,wu2025microblossom}.

\begin{figure*}[t]
    \centering
    \includegraphics[width=\linewidth]{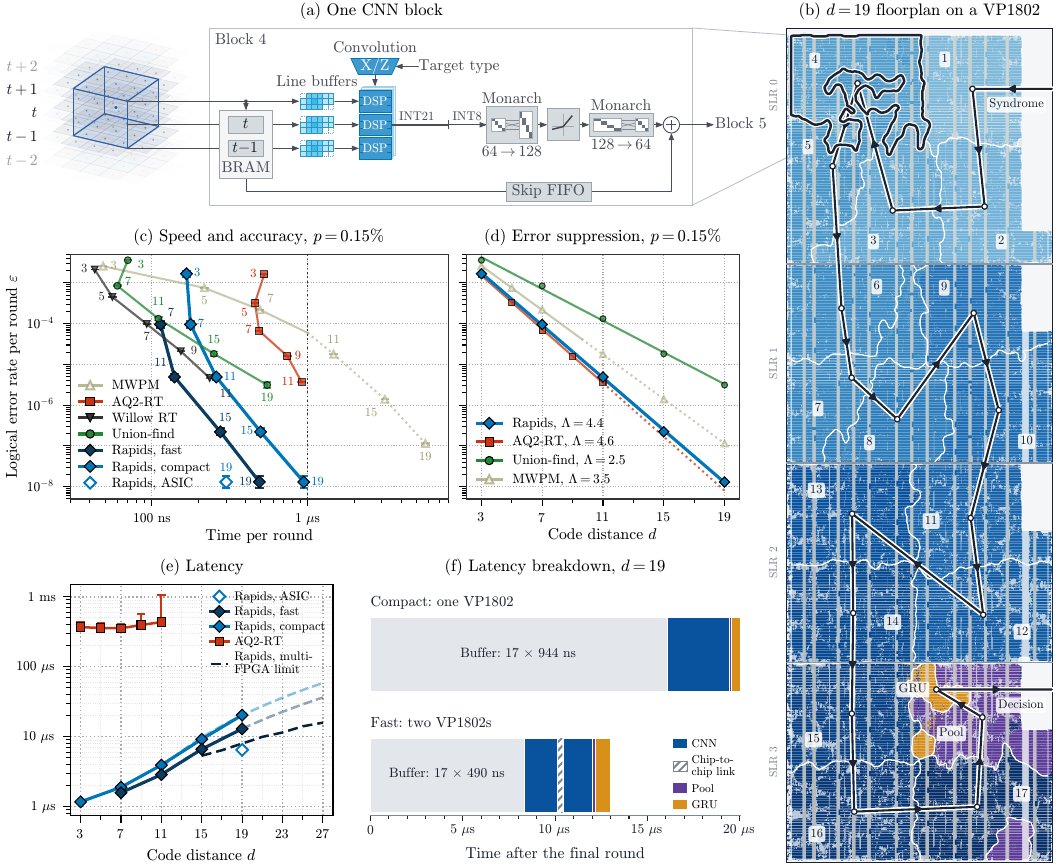}
    \caption{\textbf{Rapids on the surface code.}
    (a)~Implementation of one neural network block in hardware.
    (b)~Decoder floorplan for Rapids at $d=19$ on an AMD VP1802 FPGA.
    (c,d)~Surface-code memory experiments under SI1000 noise at $p=0.15\%$. 
    AQ2-RT~\cite{alphaqubit2}, Willow RT~\cite{google2024willow}, and union-find~\cite{barber2025realtime,liyanage2024fpga} are prior real-time decoders.
    Rapids compact and fast are two FPGA implementations of the same trained decoder: compact uses a single FPGA, and fast variants use more hardware (up to two FPGAs) to decode faster; the open diamond is an estimate of speed on an ASIC (Methods). (c)~Logical error rate against throughput, with the dotted line indicating the 1~MHz throughput required for superconducting processors. (d)~Error suppression factor $\Lambda$ (lines are dotted where a decoder does not reach 1~MHz throughput).
    (e)~Latency against code distance, extrapolated to $d=27$ (faded). The dark dashed line is an estimate of the lowest latency achievable upon scaling to arbitrarily many FPGAs (\cref{fig:limits}). We show the median and 99th percentile latency of AQ2-RT~\cite{alphaqubit2}. (f)~Breakdown of the $d=19$ latency. Doubling the hardware halves the buffer latency, but adds a small chip-to-chip communication delay.}
    \label{fig:decoder}
\end{figure*}

Neural decoders avoid this trade-off in principle.
They achieve near-optimal accuracy with a computation structure that is highly parallel and therefore, on suitable hardware, extremely fast~\cite{bausch2024learning,ataides2025neural,gu2026scalableneuraldecoderspractical,alphaqubit2}.
However, two obstacles have kept them from working in real time.
First, most neural decoders process a complete memory experiment at once~\cite{zhang2023scalable,lange2023datadriven,gu2026scalableneuraldecoderspractical}, whereas a real-time decoder must process an unbounded stream of syndrome rounds as they are measured.
Second, accurate neural decoders have run in real time only on machine-learning (ML) accelerators, which are designed for throughput rather than latency. For instance, the real-time variant of the AlphaQubit~2 decoder has latencies of up to a millisecond~\cite{alphaqubit2}, two orders of magnitude above the $10~\mu$s reaction time assumed by typical resource estimates~\cite{gidney2021factor,kivlichan2020improved,lee2021even,litinski2023ecc,gidney2025factor}.
Even setting latency aside, no neural decoder has yet kept pace with a $1~\mu$s syndrome period beyond $d=11$, either on ML accelerators~\cite{alphaqubit2} or on dedicated hardware, where demonstrations have reached at most $d=7$~\cite{overwater2022neural,zhang2023scalable,cicero2026low,yang2026realtime}.

We introduce \emph{Rapids}, a family of fast and accurate streaming neural decoders designed to run as dedicated circuits on field-programmable gate arrays (FPGAs).
We co-design each decoder, from the architecture to individual layers and training, with both the FPGA that runs it and the code it decodes (\cref{fig:architecture}).
It builds on Cascade~\cite{gu2026scalableneuraldecoderspractical}, a convolutional neural decoder with near-optimal accuracy, adding a small recurrent network that makes it a parallel-window decoder of the syndrome as it arrives.
The convolutional network uses sparse, hardware-efficient layers tailored to the arithmetic and logic resources of the FPGA.
Quantization-aware training preserves its accuracy in the integer arithmetic of the device.
For the distance-19 surface code, Rapids keeps pace with a $1~\mu$s syndrome period on a single FPGA, with $20~\mu$s of latency and 240 times fewer logical errors than existing real-time decoders~\cite{barber2025realtime}.
The decoder is also \emph{scalable}: spread over more FPGAs, it runs faster or more accurately, and it can be divided in space for codes assembled from smaller modules~\cite{higgott2026modular,wills2026concatenating}.

\section*{Results}
\label{sec:results}
\label{sec:streaming}
\phantomsection\label{sec:codesign}

\begin{figure*}[t]
    \centering
    \includegraphics[width=\linewidth]{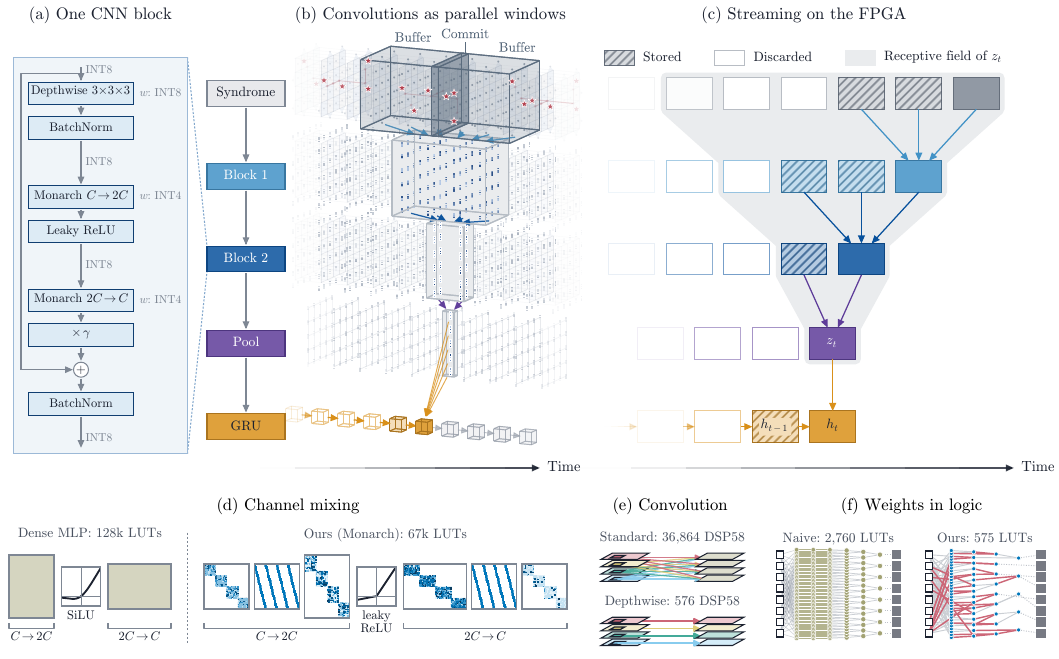}
    \caption{\textbf{Rapids as a parallel-window decoder, co-designed with the FPGA.} (a)~One CNN block, with numerical precision of the weights $w$ and activations. (b)~With $L=2$ CNN blocks, the pooled vector $z_t$ depends on a bounded window of input rounds. This vector $z_t$ can be understood as a representation of the commitment for round $t$, with buffers of size $L$ on either side. (c)~This leads to a streaming design on an FPGA: each layer computes the features of one new round from the rounds it stores in block RAM (hatched), so the layers work on different rounds at once.
    (d)~Monarch linear layers~\cite{dao2022monarch} and hardware-efficient nonlinearities sparsify each multilayer perceptron, halving its LUTs.
    (e)~Sparse depthwise convolutions further reduce hardware requirements.
    (f)~Fixed weights compiled into shift-and-add logic require 5 times fewer LUTs than general multipliers.}
    \label{fig:architecture}
\end{figure*}

A decoder receives syndromes as an unbounded stream of data. It must continually process this stream, updating its internal state to infer which errors have occurred.
Two quantities characterize its speed.
The \emph{throughput} is the number of rounds processed per second, which must match the rate of syndrome extraction.
The latency is the time between a transversal readout and the decoder's output of the corresponding correction.

Windowed decoders process a stream by dividing it into windows of rounds that overlap in time, either sequentially, as sliding windows, or concurrently, as parallel windows~\cite{dennis2002topological,skoric2023parallel,tan2023scalable,bombin2023modular}.
A sliding window commits the corrections for its earliest rounds, using its later rounds as a buffer, and passes them to the next window as a boundary condition~\cite{dennis2002topological,skoric2023parallel}.
Each window must wait for the one before it, so the throughput is limited by the time to decode a single window.
Parallel windows remove this dependence~\cite{skoric2023parallel,tan2023scalable,bombin2023modular}.
Each window has buffers on both sides and is decoded independently of its neighbors.
The throughput then grows with the number of windows decoded at once.
In what follows, we show that the convolutional decoder Cascade~\cite{gu2026scalableneuraldecoderspractical}, augmented with a small recurrent network, is a parallel-window decoder of this kind.

A convolutional neural network (CNN) applies the same small filter at every position of a grid, much as a code repeats the same check across the lattice.
Decoders built this way reach near-optimal accuracy~\cite{gu2026scalableneuraldecoderspractical}.
In a CNN decoder, each check is a site carrying a vector of $C$ features.
Each of its $L$ blocks updates the features of every site from those of its neighbors in space and time, with the same weights at every site and round (\cref{fig:architecture}(a)).
Larger values of the width $C$ and the depth $L$ make the decoder more expressive, hence more accurate, but also larger and more expensive to implement (\cref{fig:scaling}).
In their original form~\cite{gu2026scalableneuraldecoderspractical}, however, these networks decode a memory experiment of fixed length as a whole, and so cannot decode an unbounded stream of syndromes.
Their structure nonetheless already contains parallel windows.
In the surface-code decoder, each convolutional block reads the round before and the round after, so after $L$ blocks the features of round $t$ depend only on rounds $t-L$ to $t+L$ (\cref{fig:architecture}(b)).
The CNN thus computes a window of $2L+1$ rounds around every round, each independently of the others.
Replacing the original readout with a small recurrent network therefore turns the CNN into a parallel-window decoder: a logical pool sums the features of each round over the support of each logical operator, and a gated recurrent unit (GRU)~\cite{cho2014gru} accumulates these pooled vectors into a prediction of the cumulative logical flip.

The structure of the network suits an FPGA implementation in which every block has its own dedicated hardware.
Each block depends only on the output of the previous block, so the blocks can work at the same time on different rounds: while one block processes a round, the previous block is already processing the next (\cref{fig:architecture}(c)).
The blocks thus form a pipeline, through which the syndrome streams one site per clock cycle.
A recurrent layer must finish its update for one round before it can start the next, so Rapids keeps its only recurrence in the small GRU at the end; the CNN carries a bounded history of features and is pipelined across rounds.

In this implementation, the throughput does not depend on the depth $L$ of the network.
The decoder finishes rounds as fast as a single block processes one, however many blocks there are, so each added block costs hardware but does not change the throughput.
The latency, in contrast, does depend on the depth.
Each block's convolution reads one round into the future, so the output for a round depends on rounds up to $L$ ahead of it.
Once the final round is read out, these convolutions still read $L$ rounds that never arrive; as at the edge of an image, the decoder fills them with zeros.
The latency is this buffer of zeros plus the time for a single round to traverse the pipeline.

\begin{figure*}[t]
    \centering
    \includegraphics[width=\linewidth]{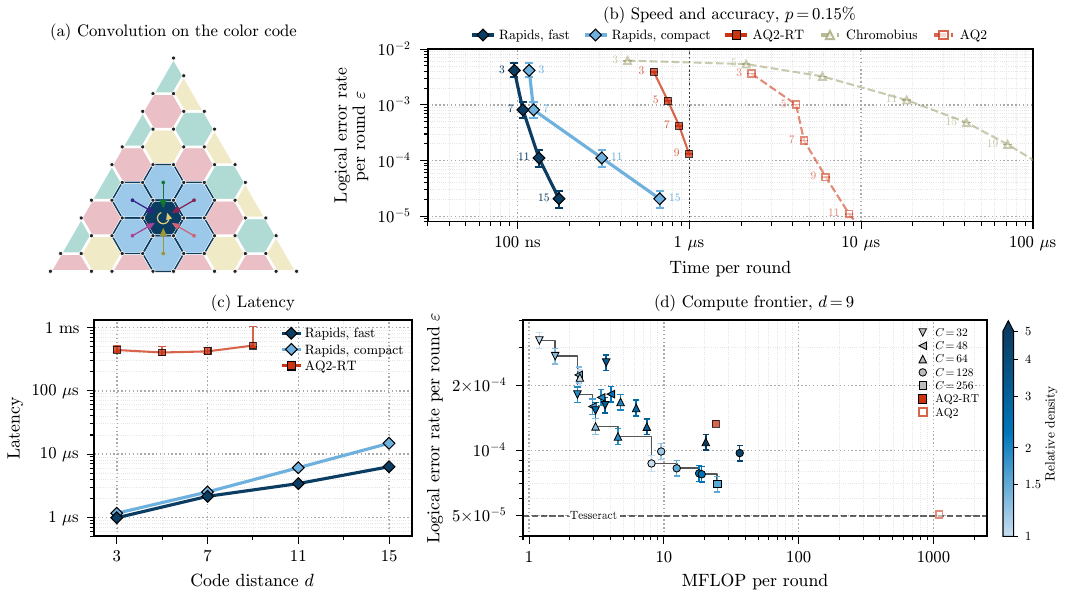}
    \caption{\textbf{Color-code.} (a)~$d=9$ color code: each tile holds an $X$ and a $Z$ check, and dots mark the data qubits. Each convolution computes a tile (dark) from itself and its six neighbors (light). (b)~Logical error rate against the time per round, with Rapids compact and fast as in \cref{fig:decoder}.
    Chromobius, AQ2-RT, and AQ2 show values from Ref.~\cite{alphaqubit2}.
    (c)~Latency against code distance, with AQ2-RT as in \cref{fig:decoder}(e).
    (d)~Logical error rate against the floating-point operations (FLOPs) per round at $d=9$ for models with different channel widths $C$ and model densities (see Methods).}
    \label{fig:color_multi}
\end{figure*}

The network must also fit on the real hardware.
An FPGA provides three kinds of resource: lookup tables (LUTs), which are small units of configurable logic; digital signal processing slices (DSPs), dedicated units for integer multiplication; and block RAM (BRAM) for storage.
The original $d=19$ network of Ref.~\cite{gu2026scalableneuraldecoderspractical} would need about 50 times the DSP slices of an AMD Versal Premium VP1802 for its convolutions alone.
To close this gap, we build the network using only operations that map cheaply onto FPGA resources. This trades some of the accuracy of an unconstrained network for a large saving in hardware, and at a fixed hardware budget such sparse networks are in fact the more accurate choice (\cref{fig:color_multi}(d)).

A standard $3\times3\times3$ convolution mixes every channel of every neighbor, requiring $27C^2$ products per site, which dominates the cost of a convolutional network.
We use depthwise convolutions instead, which convolve each channel separately, requiring only $27C$ products~\cite{howard2017mobilenets,liu2022convnet} (\cref{fig:architecture}(e)).
At the widths we use, depthwise convolution uses 32 to 128 times fewer DSP slices than standard dense convolution.
However, a depthwise convolution never mixes channels, so it is usually followed by a small multilayer perceptron (MLP) that does.
To make this mixing cheap as well, we replace each dense map of the MLP by a Monarch matrix~\cite{dao2022monarch}: two block-diagonal steps joined by a fixed permutation of the channels. Together, this still connects every input channel to every output channel, with 3 to 5 times fewer products (\cref{fig:architecture}(d)).

While GPUs are built around floating-point arithmetic, FPGAs more naturally accommodate integer arithmetic.
We therefore quantize the depthwise convolutions and Monarch layers to 8-bit and 4-bit integers. To preserve decoder accuracy, we train the network with this quantization in place (\cref{fig:architecture}(a))~\cite{bengio2013ste,esser2020lsq}. Since the Monarch weights are 4-bit integers, each product in a matrix multiplication reduces to a few shifts and additions: $5x$, for instance, is $4x+x=(x\ll 2)+x$. We compile each Monarch step into such a network directly in the configurable logic~\cite{sun2025da4ml} (\cref{fig:architecture}(f)).
The MLPs then need no multipliers, which is a dramatic savings. Implemented on DSP slices, the MLPs of our $d=19$ decoder alone would need more than twice the capacity of a VP1802, but compiled as logic, they take about a third of its LUTs.
For further savings, we use a leaky ReLU with a negative slope of $1/8$ as the nonlinearity, which needs only a comparison and a shift, whereas a smooth activation would need additional hardware (e.g., a multiplier or a lookup table per channel).

\begin{figure*}[htbp!]
    \centering
    \includegraphics[width=\linewidth]{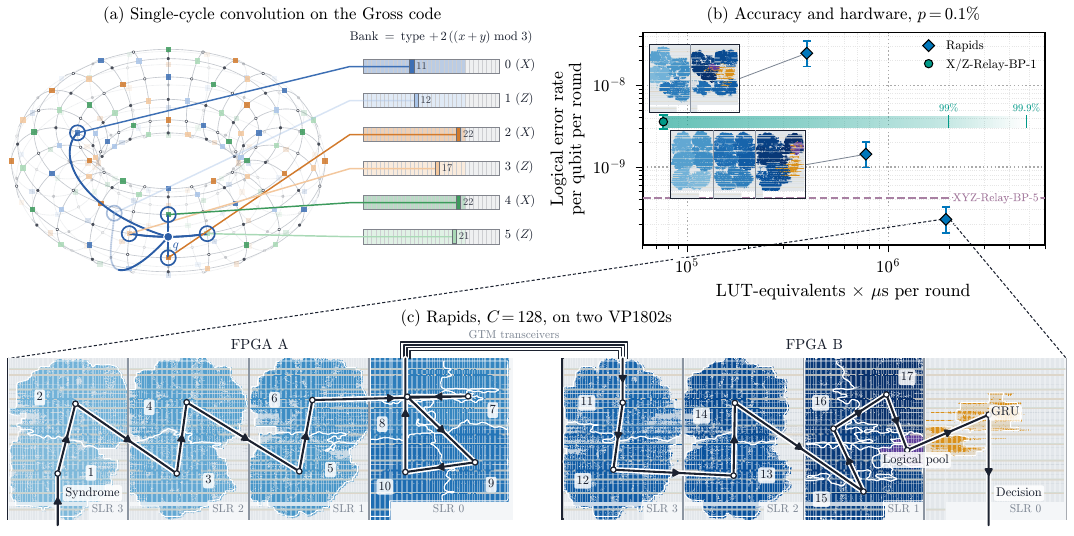}
    \caption{\textbf{Gross code.} (a)~The $\llbracket144,12,12\rrbracket$ code on a torus. Squares are checks and dots are data qubits. Colors mark the six memory banks; the six checks of a data qubit $q$ lie in different banks, so a convolution reads all six in one clock cycle. (b)~Logical error rate per logical qubit per round at $p=0.1\%$ as a function of the hardware space-time cost of the decoder. The dashed line shows offline (i.e., not real-time) joint-basis floating-point Relay decoding~\cite{muller2024relax}. X/Z-Relay-BP-1 is the real-time int4 variant of Relay proposed in Ref.~\cite{maurer2025relay}. We show the median, 99\%, and 99.9\% cost of Relay, whose runtime depends on the syndrome. (c)~The most accurate decoder (which used $C=128$ channels) laid out on two VP1802s at $0.33~\mu$s per round, joined by a stream of activations over transceiver links (Methods).}
    \label{fig:other_codes}
\end{figure*}

The local, feedforward structure of the CNN also lets the decoder scale with hardware.
The circuit is thus parallel over the channels of each site, and pipelined over sites within a block and over rounds across blocks.
These scale independently: processing more sites in parallel improves the throughput and latency, and a wider network improves the accuracy without reducing the throughput.
Importantly, because each convolutional block reads only the output of the block before it, over a bounded neighborhood, the pipeline can be cut at any block boundary and its stages placed on different devices (\cref{fig:scaling}).
The cut carries a single stream of activations consisting of $C$ INT8 channels per site. For instance, our fast $d=19$ surface-code build has a stream rate of 418~Gb/s, well within the reach of standard optical transceivers.

Decoders based on belief propagation or on matching cannot be divided this way.
Although their throughput can be multiplied by assigning different windows to different devices, this faces a fundamental scaling barrier: each window must fit on a single device. This owes to the structure of their computation: rather than passing information forward through a fixed sequence of layers, they iterate over the whole decoding graph. Splitting a window across devices therefore means partitioning its decoding graph, with chip-to-chip links carrying information across the edges that cross the partition. Since any partition cuts many edges, and the decoder sends information across them in every iteration, the delay of these links would make decoding prohibitively slow.
This is clearest for belief propagation, which passes messages along every edge of the Tanner graph for tens to hundreds of iterations per window. Split across two devices, every iteration would wait for a round trip over the link: at about $0.5~\mu$s per round trip, a window of a hundred iterations would take at least $50~\mu$s, too slow for superconducting hardware.
The real-time variant of the Relay-BP decoder already needs over half of an FPGA to decode the Gross code in just one basis~\cite{maurer2025relay}. This leaves no clear path to scale towards joint-basis decoding or to larger codes.

We implement Rapids for three codes: the surface code, the Gross code, and the color code. For both the surface and color code, we evaluate our decoders under SI1000 noise, and for the Gross code, we use standard depolarizing noise.
We place and route each decoder on AMD Versal Premium VP1802 FPGAs and report its speed, accuracy, and hardware footprint.

\backtopic{Surface code}
We begin with the rotated surface code.
At $d=19$, a single FPGA suffices to run a decoder that processes over one million syndrome rounds per second and is significantly more accurate than matching (\cref{fig:decoder}(c,d)).
Its logical error rate at $p=0.15\%$ is 240 times below that of union-find, which is (to our knowledge) the only decoder previously shown to run in real time at this distance~\cite{barber2025realtime,liyanage2024fpga}, and nearly 10 times below that of MWPM.
This advantage grows with the code distance: Rapids suppresses errors by a factor of $\Lambda=4.4$ for every two units of distance, compared with 3.5 for MWPM and about 2.5 for union-find (\cref{fig:decoder}(d)).
Importantly, this accuracy corresponds to a saving in qubits: union-find would need about $d=31$ to match the logical error rate of Rapids at $d=19$, so Rapids reaches the same error rate with 62\% fewer physical qubits (721 instead of 1{,}921 per logical qubit).
Furthermore, as we show in \cref{fig:extrapolation}(b), the decoder accuracy remains stable on long memory experiments.

Beyond accuracy, decision latency is a key performance metric: the run time of most compiled fault-tolerant algorithms depends strongly on this latency~\cite{gidney2021factor}.
Our compact (single-FPGA) implementation of Rapids has a decision latency of $20.36~\mu$s at $d=19$. Most of this latency is the buffer (\cref{fig:decoder}(f)), with a small fixed addition of $4.3~\mu$s for the pipeline of CNN, pool and GRU. The buffer time scales inversely with the throughput, so the most effective way to improve the latency is to increase the throughput.
It may appear that the throughput is fundamentally limited by the clock rate of the FPGA though: our compact implementation of Rapids processes one site per clock cycle, so a round takes $(d+1)^2$ cycles. Since large designs on current FPGAs can reach at most 500~MHz, this limits the throughput to $500/(d+1)^2$ syndrome rounds per microsecond.
However, this is not fundamental, because decoding can also be \emph{parallelized spatially}.
Specifically, the fast variant of Rapids processes two sites per clock cycle, which doubles the throughput, at the cost of double the FPGA hardware. It runs at $0.49~\mu$s per round, which nearly halves the latency to $13~\mu$s (\cref{fig:decoder}(f)).
As this example illustrates, throughput scales linearly with hardware: $N$ FPGAs can process $N$ times as many sites per clock cycle. Although this method of scaling allows throughput to increase indefinitely, as we show in \cref{fig:decoder}(e), latency only falls until the chip-to-chip communication delay outweighs the shorter buffer, which sets a floor of $\sim 8~\mu$s at $d=19$.

The same circuit also maps to custom silicon.
An estimate in a 7~nm process node gives a throughput of $0.30~\mu$s per round and a latency of $6.5~\mu$s (Methods). This ASIC design uses considerably less hardware area and power than our FPGA design: it draws about 10~W of power (against 250~W for an FPGA) with an area of 7.4~mm$^2$.

\backtopic{Gross code}
Because our design does not require a planar layout, it extends to quantum low-density parity-check codes. We demonstrate this for the $\llbracket 144,12,12\rrbracket$ bivariate-bicycle (BB) code~\cite{bravyi2024high}, known as the Gross code.
Since its checks act on qubits that are far apart in any planar arrangement, a geometrically local convolution is no longer possible, and must be replaced by a long-range convolution that connects each check to its six neighbors in the Tanner graph (\cref{fig:other_codes}(a)).
As these neighbors are spread across the code, they must be read from memories (BRAMs) that hold the entire round. However, a BRAM returns only one word per read port in each clock cycle, so when two neighbors are stored in the same memory, they must be read sequentially, halving the decoder's throughput.
We avoid such conflicts by dividing the storage into six memory banks, assigned by a simple rule so that the six neighbors of any node lie in different banks (\cref{fig:other_codes}(a); Methods); the decoder then processes one check per clock cycle.

We have shown that additional hardware can be used to improve decoding speed; we now show that it can also improve accuracy.
We train three decoders with different channel widths $C$, and build each on FPGAs.
We quantify the cost of the decoder by its hardware cost (measured by LUT-equivalents; see Methods) multiplied by its time per round.
As shown in \cref{fig:other_codes}(b), by scaling up the hardware fivefold, the logical error rate can be decreased by a factor of 100.
The same applies to power, from 28 to 111~W at one round per microsecond (\cref{tab:resources}), so four times the power buys a hundred times fewer logical errors.
This costs almost nothing in speed: a round takes 144 clock cycles at every width, so the time per round only grows from 0.30 to $0.33~\mu$s.
Our most accurate decoder (\cref{fig:other_codes}(c)) has a decision latency of $8.0~\mu$s.
At $p=0.1\%$ it is 16 times more accurate than the real-time Relay-BP-1 decoder~\cite{maurer2025relay}, exceeding even the accuracy of joint-basis Relay-BP~\cite{muller2024relax} (which has no real-time hardware implementation).

\begin{figure*}[t]
    \centering
    \includegraphics[width=\linewidth]{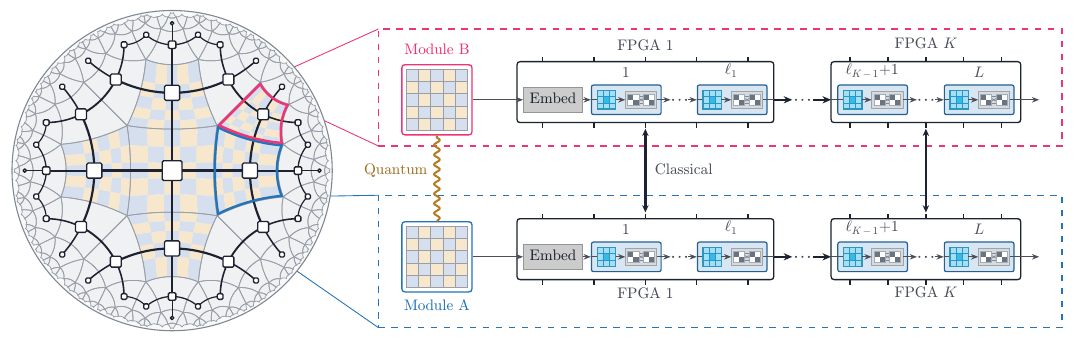}
    \caption{\textbf{Splitting the decoder across FPGAs and across modules.} Left: a modular hyperbolic surface code~\cite{higgott2026modular}, drawn schematically as a $\{4,5\}$ tiling of the Poincar\'e disk. Each face is one planar module, with its own decoder (white squares); a link joins the decoders of two modules that share a seam. Right: two adjacent modules, A and B (outlined). The readout of each module enters its own decoder, whose $L$ blocks are cut at block boundaries over $K$ FPGAs. The site stream, $C$ INT8 channels per site, leaves each FPGA once and enters the next. A quantum coupler joins the qubits of A and B, and a classical link joins FPGA $i$ of A to FPGA $i$ of B, carrying the sites at the seam that the convolutions on these FPGAs need.}
    \label{fig:multi_fpga}
\end{figure*}

\backtopic{Color code}
To further illustrate the generality of our framework, we also implement a decoder for the superdense color code~\cite{bombin2006topological}, which differs in structure from the surface code but still has a local convolution, over each tile and its six neighbors (\cref{fig:color_multi}(a)).
Color codes support transversal Clifford gates and underlie magic state cultivation~\cite{gidney2024magic}, but they are harder to decode: matching-based decoders such as Chromobius~\cite{gidney2023chromobius} give up much of the code's accuracy, and more accurate search decoders such as Tesseract~\cite{shutty2025tesseract} are far too slow for real time.
Neural decoders close this gap: AlphaQubit~2 (AQ2) reaches an error rate about 500 times below that of Chromobius at $d=15$~\cite{alphaqubit2}.
Its real-time variant, AQ2-RT, however, keeps pace with a $1~\mu$s round only up to $d=9$, and its decision latency can be up to 1~ms.

At $p=0.15\%$ and $d=15$ our decoder is over 20 times more accurate than Chromobius (\cref{fig:color_multi}(b)).
The compact builds fit on a single FPGA and keep pace with the syndrome up to $d=15$, and the fast builds, which span two FPGAs at $d=15$, take $0.17~\mu$s per round and have a decision latency of $6.3~\mu$s.
Moreover, as we show below, the accuracy can be improved with more decoder hardware.

We train a series of models at $d=9$ with widths $C$ from 32 to 256 and a range of densities.
Two choices set the density of a model: its convolution, which ranges from depthwise, where each channel is convolved separately, to grouped, where each output channel reads several input channels; and its channel mixing, which ranges from a Monarch matrix of many blocks to a dense matrix.
Its density is its number of weights relative to the sparsest model of the same width (Methods).
This serves as a test of the central choice of our decoder co-design, namely its \emph{sparsity} (\cref{fig:architecture}(d,e)).
We measure the cost of each model by its floating-point operations (FLOPs) per round: on an FPGA, every operation of the network has its own circuit, so at a fixed throughput the operations per round set the hardware that the decoder needs.
As we illustrate in \Cref{fig:color_multi}(d), sparse models are preferable.
A likely reason for this is that, all else being equal, sparser networks allow larger channel widths $C$, and this width sets how much information each site can hold.
For instance, although they both require around 8 million FLOPs per round, the sparsest model of width 128 makes 1.5 times fewer errors than a model of width 64 that is three times as dense.
This efficiency means our models can recover the performance of AQ2-RT with about an eighth of its FLOPs per round. 
With the same FLOP budget, they are twice as accurate, which suggests that more hardware would recover the remaining accuracy gap between Rapids and AQ2-RT.

\section*{Discussion}
\label{sec:discussion}
\phantomsection\label{sec:modular}

We have shown that a neural decoder co-designed with an FPGA can be both fast enough to keep pace with superconducting hardware, and more accurate than state-of-the-art real-time decoders~\cite{barber2025realtime,liyanage2024fpga,maurer2025relay}. 

Just as its blocks are divided over FPGAs, the decoder can also be divided along space, which suits large codes built from small planar modules joined by a few couplers, such as a modular hyperbolic surface code~\cite{higgott2026modular} (\cref{fig:multi_fpga}).
Each module then has its own decoder, and the decoders need to exchange very little information.
A convolution reads only the neighbors of a site, so a block next to a seam needs from the adjacent module only the sites within one step of the seam: a strip one site wide, in each round and for each block.
The communication traffic therefore grows with the length of the seam, not with the area of the modules.
A decoder for an arbitrarily large modular code is then many identical small decoders joined by short, fixed links: its hardware grows in proportion to the number of modules, and its throughput stays that of a single module.
A global decoder based on matching or belief propagation, by contrast, would have to reconcile its solution across every seam.

Compiled fault-tolerant algorithms of practical interest use thousands of logical qubits~\cite{gidney2021factor}, each with its own decoder running continuously.
At $d=19$, one VP1802 per logical qubit would amount to thousands of FPGAs and several hundred kilowatts, the equivalent of tens of server racks.
High-rate codes and custom silicon could each reduce this by an order of magnitude: our largest Gross-code decoder draws about 9~W per logical qubit, against about 250~W for a single surface-code logical qubit at $d=19$, and our ASIC layout of the $d=19$ decoder about 10~W in all.
Together, if the savings of custom silicon carry over to high-rate codes, the two would bring the power to about 0.4~W per logical qubit, or under a kilowatt for thousands of logical qubits.
Regardless, at this scale, decoding hardware becomes a nontrivial consideration, whose power and cost should enter the design of fault-tolerant architectures alongside those of the qubits.

\begin{acknowledgments}
We thank J. Pablo Bonilla Ataides, Hengyun (Harry) Zhou, and Marco Venere for helpful discussions.
We acknowledge support from Anthropic under the AI for Science program.
We thank the AMD University Program for donating the Vivado Design Suite licenses used in this work.
This research used resources of the National Energy Research Scientific Computing Center (NERSC), a Department of Energy Office of Science User Facility under Contract No. DE-AC02-05CH11231 using NERSC award DDR-ERCAP0038713.
The computations in this paper were run in part on the FASRC Cannon cluster supported by the FAS Division of Science Research Computing Group at Harvard University.
A.G. and S.F.Y. acknowledge the NSF via the ORAQL NQVL QSTD.
\end{acknowledgments}

\section*{Author contributions}
A.G. conceived the project and designed the decoder architecture. A.G. and C.J. designed the FPGA circuits. C.J. implemented the decoders as circuits on FPGAs and characterized their performance. S.F.Y. supervised the project. All authors wrote and revised the manuscript.

\section*{Competing interests}
S.F.Y.'s spouse is a co-founder and shareholder of QuEra Computing. The other authors declare no competing interests.

\renewcommand*{\bibfont}{\footnotesize}
\bibliography{refs}

\clearpage
\setcounter{figure}{0}
\setcounter{table}{0}
\renewcommand{\figurename}{Extended Data Fig.}
\renewcommand{\tablename}{Extended Data Table}
\renewcommand{\thetable}{\arabic{table}}
\renewcommand{\theHfigure}{ED\arabic{figure}}
\renewcommand{\theHtable}{ED\arabic{table}}

\onecolumngrid
\section*{Methods}

\subsection*{Training}
\phantomsection\label{meth:training}

\begin{table*}[b]
    \centering
    \footnotesize
    \setlength\tabcolsep{4.5pt}
    \renewcommand\arraystretch{0.95}
    \setlength\aboverulesep{0.25ex}\setlength\belowrulesep{0.35ex}
    \begin{tabular}{@{}llrrrrrrcrrrrrrr@{}}
\toprule
 & & & & & & Clock & Round & & \multicolumn{3}{c}{Share of FPGA (\%)} & \multicolumn{3}{c}{Latency ($\mu$s)} & Power \\
\cmidrule(lr){10-12}\cmidrule(l){13-15}
Code & Build & $d$ & $C$ & $n_b$ & $L$ & (MHz) & ($\mu$s) & FPGA & LUT & FF & DSP & Buffer & Pipeline & Total & (W) \\
\midrule
\multirow{12}{*}[-\dimexpr 9\aboverulesep/2+9\belowrulesep/2\relax]{Surface} & \multirow{5}{*}[-\dimexpr 4\aboverulesep/2+4\belowrulesep/2\relax]{Compact} & 3 & 16 & 4 & 3 & 606.1 & 0.167 &  & 1.4 & 1.7 & 6.4 & 0.50 & 0.67 & 1.17 & 12 \\
\rulecolor{black!15}\cmidrule[0.3pt](l){3-16}
 &  & 7 & 24 & 4 & 5 & 571.4 & 0.177 &  & 3.8 & 3.9 & 11.8 & 0.88 & 1.01 & 1.90 & 14 \\
\rulecolor{black!15}\cmidrule[0.3pt](l){3-16}
 &  & 11 & 32 & 4 & 9 & 555.6 & 0.259 &  & 9.6 & 9.8 & 24.0 & 2.33 & 1.61 & 3.94 & 26 \\
\rulecolor{black!15}\cmidrule[0.3pt](l){3-16}
 &  & 15 & 48 & 4 & 13 & 512.8 & 0.499 &  & 25.4 & 22.8 & 44.9 & 6.49 & 2.72 & 9.21 & 82 \\
\rulecolor{black!15}\cmidrule[0.3pt](l){3-16}
 &  & 19 & 64 & 4 & 17 & 423.7 & 0.944 &  & 49.7 & 52.2 & 77.5 & 16.05 & 4.31 & 20.36 & 246 \\
\rulecolor{black!45}\cmidrule[0.4pt]{2-16}
 & \multirow{6}{*}[-\dimexpr 3\aboverulesep/2+3\belowrulesep/2\relax]{Fast} & 7 & 24 & 4 & 5 & 537.6 & 0.113 &  & 7.0 & 7.2 & 24.4 & 0.57 & 1.02 & 1.58 & 14 \\
\rulecolor{black!15}\cmidrule[0.3pt](l){3-16}
 &  & 11 & 32 & 4 & 9 & 515.5 & 0.140 &  & 17.7 & 17.2 & 45.5 & 1.26 & 1.62 & 2.88 & 24 \\
\rulecolor{black!15}\cmidrule[0.3pt](l){3-16}
 &  & \multirow{2}{*}{15} & \multirow{2}{*}{48} & \multirow{2}{*}{4} & \multirow{2}{*}{13} & \multirow{2}{*}{465.1} & \multirow{2}{*}{0.275} & 1/2 & 24.0 & 21.2 & 46.8 & \multirow{2}{*}{3.58} & \multirow{2}{*}{3.05} & \multirow{2}{*}{6.63} & \multirow{2}{*}{82} \\
 &  &  &  &  &  &  &  & 2/2 & 22.6 & 19.9 & 43.1 &  &  &  &  \\
\rulecolor{black!15}\cmidrule[0.3pt](l){3-16}
 &  & \multirow{2}{*}{19} & \multirow{2}{*}{64} & \multirow{2}{*}{4} & \multirow{2}{*}{17} & \multirow{2}{*}{408.2} & \multirow{2}{*}{0.490} & 1/2 & 50.4 & 45.8 & 77.3 & \multirow{2}{*}{8.33} & \multirow{2}{*}{4.65} & \multirow{2}{*}{12.98} & \multirow{2}{*}{216} \\
 &  &  &  &  &  &  &  & 2/2 & 49.0 & 42.6 & 74.8 &  &  &  &  \\
\rulecolor{black!45}\cmidrule[0.4pt]{2-16}
 & ASIC & 19 & 64 & 4 & 17 & 1333.3 & 0.300 &  & \multicolumn{3}{c}{7.4 mm$^2$} & 5.10 & 1.40 & 6.50 & 10 \\
\rulecolor{black!45}\midrule[0.4pt]
\multirow{9}{*}[-\dimexpr 7\aboverulesep/2+7\belowrulesep/2\relax]{Color} & \multirow{4}{*}[-\dimexpr 3\aboverulesep/2+3\belowrulesep/2\relax]{Compact} & 3 & 16 & 4 & 3 & 598.8 & 0.117 &  & 1.4 & 1.6 & 13.7 & 0.35 & 0.81 & 1.16 & 12 \\
\rulecolor{black!15}\cmidrule[0.3pt](l){3-16}
 &  & 7 & 40 & 5 & 7 & 588.2 & 0.124 &  & 7.1 & 6.7 & 20.0 & 0.87 & 1.64 & 2.51 & 15 \\
\rulecolor{black!15}\cmidrule[0.3pt](l){3-16}
 &  & 11 & 72 & 8 & 11 & 581.4 & 0.310 &  & 18.5 & 17.0 & 21.7 & 3.41 & 2.67 & 6.07 & 29 \\
\rulecolor{black!15}\cmidrule[0.3pt](l){3-16}
 &  & 15 & 128 & 8 & 15 & 500.0 & 0.672 &  & 60.1 & 54.6 & 47.0 & 10.08 & 4.71 & 14.79 & 109 \\
\rulecolor{black!45}\cmidrule[0.4pt]{2-16}
 & \multirow{5}{*}[-\dimexpr 3\aboverulesep/2+3\belowrulesep/2\relax]{Fast} & 3 & 16 & 4 & 3 & 584.8 & 0.096 &  & 1.9 & 2.3 & 15.2 & 0.29 & 0.69 & 0.98 & 12 \\
\rulecolor{black!15}\cmidrule[0.3pt](l){3-16}
 &  & 7 & 40 & 5 & 7 & 520.8 & 0.108 &  & 11.3 & 12.2 & 30.5 & 0.75 & 1.39 & 2.14 & 15 \\
\rulecolor{black!15}\cmidrule[0.3pt](l){3-16}
 &  & 11 & 72 & 8 & 11 & 555.6 & 0.133 &  & 32.9 & 33.4 & 56.5 & 1.47 & 1.95 & 3.41 & 30 \\
\rulecolor{black!15}\cmidrule[0.3pt](l){3-16}
 &  & \multirow{2}{*}{15} & \multirow{2}{*}{128} & \multirow{2}{*}{8} & \multirow{2}{*}{15} & \multirow{2}{*}{487.8} & \multirow{2}{*}{0.174} & 1/2 & 54.4 & 57.6 & 57.1 & \multirow{2}{*}{2.61} & \multirow{2}{*}{3.68} & \multirow{2}{*}{6.30} & \multirow{2}{*}{110} \\
 &  &  &  &  &  &  &  & 2/2 & 55.0 & 53.3 & 60.1 &  &  &  &  \\
\rulecolor{black!45}\midrule[0.4pt]
\multirow{4}{*}[-\dimexpr 2\aboverulesep/2+2\belowrulesep/2\relax]{Gross} & \multirow{4}{*}[-\dimexpr 2\aboverulesep/2+2\belowrulesep/2\relax]{} & 12 & 32 & 4 & 8 & 483.1 & 0.298 &  & 14.2 & 12.9 & 22.0 & 2.38 & 3.83 & 6.21 & 28 \\
\rulecolor{black!15}\cmidrule[0.3pt](l){3-16}
 &  & 12 & 64 & 8 & 8 & 454.5 & 0.317 &  & 27.4 & 26.0 & 39.6 & 2.53 & 4.23 & 6.76 & 44 \\
\rulecolor{black!15}\cmidrule[0.3pt](l){3-16}
 &  & \multirow{2}{*}{12} & \multirow{2}{*}{128} & \multirow{2}{*}{8} & \multirow{2}{*}{8} & \multirow{2}{*}{434.8} & \multirow{2}{*}{0.331} & 1/2 & 51.8 & 48.8 & 40.1 & \multirow{2}{*}{2.65} & \multirow{2}{*}{5.37} & \multirow{2}{*}{8.02} & \multirow{2}{*}{111} \\
 &  &  &  &  &  &  &  & 2/2 & 36.2 & 34.6 & 34.7 &  &  &  &  \\
\rulecolor{black}\bottomrule
\end{tabular}

    \caption{\textbf{Routed builds.} The decision latency includes $\sim 0.25\mu$s of chip-to-chip communication delay (Methods). We estimate power by simulating the switching activity of the decoder circuit for SI1000 syndromes at $p=0.15\%$, assuming one round of syndromes every $1~\mu$s. The `compact' and `fast' are simply different FPGA builds of the same trained decoders.}
    \label[edtable]{tab:resources}
\end{table*}

\begin{figure*}[tbp]
    \centering
    \includegraphics[width=\linewidth]{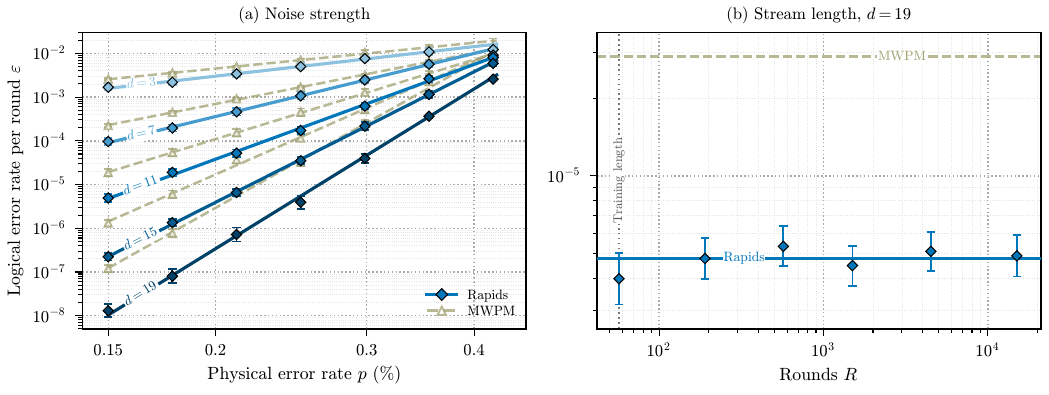}
    \caption{\textbf{Extrapolation in physical error rate and experiment length.} (a)~Logical error rate per round against the physical error rate on 120-round experiments, $d=3$ to 19. Each model is trained at a single $p$, $0.25\%$ for $d=3$ to 15 and $0.30\%$ for $d=19$. (b)~The $d=19$ decoder at $p=0.25\%$, evaluated from the CNN training length $R=3d=57$ (dotted) to $R=15{,}000$ rounds. Error bars are 95\% Wilson intervals.}
    \label[edfigure]{fig:extrapolation}
\end{figure*}

Every decoder is trained on memory experiments in the $Z$ basis, simulated with Stim~\cite{gidney2021stim}.
The surface-code decoders use the memory circuit of Ref.~\cite{alphaqubit2}, and the color-code decoders its superdense color-code circuit, both under SI1000 noise of strength $p$~\cite{gidney2022benchmarking}.
The BB decoders use standard uniform circuit-level noise of strength $p$ on the depth-8 syndrome cycle of Ref.~\cite{bravyi2024high}.
As in Ref.~\cite{gu2026scalableneuraldecoderspractical}, training begins at a low physical error rate $p$ until the training error falls to $2\%$; $p$ then rises linearly to the target.
The loss is the binary cross-entropy of the final logits, averaged over the logicals.
Each decoder is trained at a single noise level and experiment length ($R=3d$); as shown in \cref{fig:extrapolation}(a,b), the decoders extrapolate well to lower $p$ and larger $R$.

All matrix-valued parameters are optimized with NorMuon~\cite{jordan2024muon,li2025normuon}, and all other parameters with Lion~\cite{chen2023symbolic}.
We use a learning rate of $5 \times 10^{-4}$ for NorMuon and $10^{-4}$ for Lion, a warmup-stable-decay schedule that decays over the last 10\% of training steps, and a decoupled weight decay of $10^{-3}$.

\subsection*{FPGA circuits}
\phantomsection\label{meth:circuits}

All builds use Vivado 2025.2.1 and the part \texttt{xcvp1802-vsva5601-3HP-e-S}~\cite{amd_ug1582}.
For the surface code, the $3\times3$ convolution reads the $(d+1)\times(d+1)$ stabilizer grid in rounds $t-1$, $t$, and $t+1$.
Sites arrive in row-major order, one per clock cycle (two in the fast builds), so line buffers gather the spatial neighbors, block RAM supplies the preceding rounds, and coordinate and temporal masks isolate the edges of the grid and consecutive shots.
The depthwise convolution of a channel is a cascade of nine DSP58 slices, each adding three products to the partial sum of the slice before it, and a residual DSP58 combines the branches and applies the folded normalization.
The compact builds carry a pause input that holds the stream until each syndrome round arrives.
The logical pool and the GRU receive far fewer inputs per round than the CNN, so they are time-multiplexed onto a few DSP slices and floating-point engines.
The round time is therefore set by the arrival of the $(d+1)^2$ sites, except at $d\leq7$, where the GRU sets it.
The color builds follow the same design, processing one site every four clock cycles (compact) or every cycle (fast).

In the BB code, the checks lie on an $\ell\times m$ torus, with $\ell=12$ and $m=6$ for the $\llbracket144,12,12\rrbracket$ code.
A block comprises a check-to-data layer and a data-to-check layer, with a residual around the pair, and the logical pool applies a further check-to-data layer and sums over each basis logical.
Each node has a type, $X$ or $Z$ for a check and $L$ or $R$ for a data qubit, and a position $(x,y)$ on the torus, and it has three neighbors of each type, which have three different values of $(x+y) \bmod 3$.
Each node is therefore stored in bank $t+2 \cdot ((x+y) \bmod 3)$, where $t$ is 0 or 1 according to its type, so the six neighbors of any node lie in different banks.
Each layer buffers four rounds of its input in these banks, so all twelve inputs from two rounds arrive in one cycle (\cref{fig:other_codes}(a)).
The sites of a round stream in a folded column order, which brings every input neighbor of an output site within a short, fixed span.
The GRU updates the 12 logicals in turn, and the controller starts a round every 144 cycles, one per check.

\begin{figure*}[btp]
    \centering
    \includegraphics[width=\linewidth]{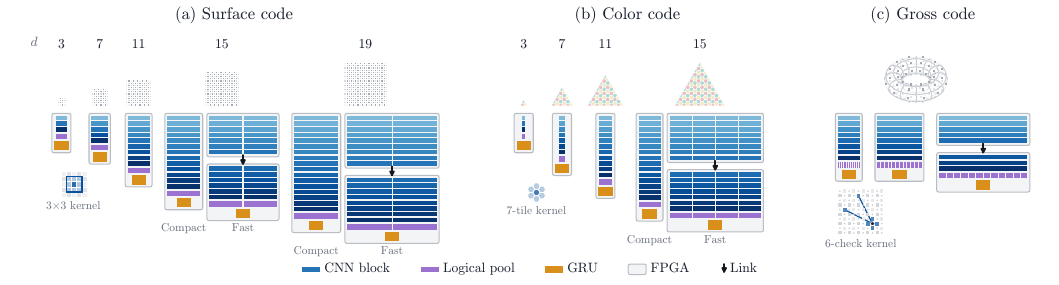}
    \caption{\textbf{Decoder implementations.} (a--c)~The decoders of the surface, color and Gross codes of \cref{tab:resources}: each runs from the syndrome of one round through $L$ CNN blocks to the logical pool and the GRU, and the width of each block is proportional to the hardware it occupies, its $C$ channels times the sites it processes per clock cycle. Fast builds process several sites per cycle in parallel lanes, and a build too large for one FPGA is cut at a block boundary and joined by a chip-to-chip link.}
    \label[edfigure]{fig:scaling}
\end{figure*}

\subsection*{Chip-to-chip links}
\phantomsection\label{si:links}

A cut through the CNN carries $8C$ bits per site and a few marker bits. This results in a communication rate of 217~Gb/s for a cut of the compact $d=19$ surface code decoder, and 418~Gb/s for the fast variant.
One eight-lane QSFP-DD connection of the VP1802's GTM transceivers carries about 217~Gb/s with Aurora 64B/66B at 28~Gb/s per lane~\cite{amd_pg074,amd_ug1582}, so the cut of the fast $d=19$ build uses two connections at 29~Gb/s per lane.
In RTL simulation, a crossing takes about $0.25~\mu$s, and a fit to these simulations gives $0.29~\mu$s for the link of the fast $d=19$ build. The 500~Gb/s cut of the fast color build at $d=15$ and the 445~Gb/s cut of the $C=128$ Gross build need two Aurora channels.

\FloatBarrier
\subsection*{ASIC estimate}
\phantomsection\label{meth:hardware}

\begin{figure*}[htbp]
    \centering
    \includegraphics[width=\linewidth]{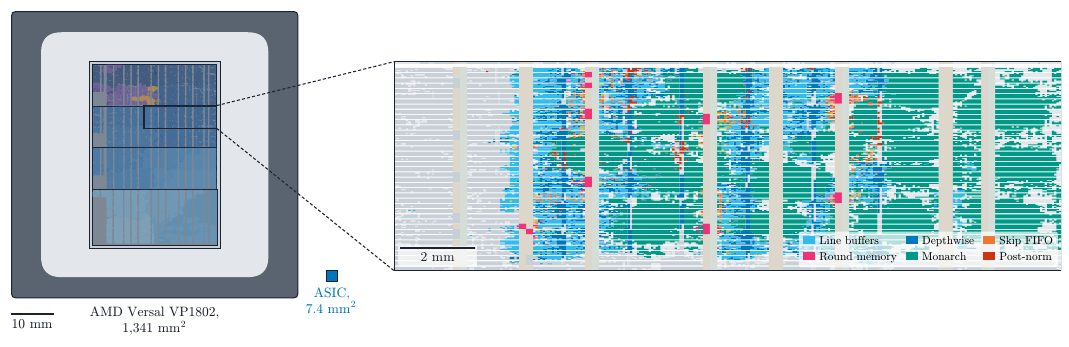}
    \caption{\textbf{The silicon of the decoder (to scale).} Left: the VP1802 in its package~\cite{amd_am013} with the $d=19$ decoder of \cref{fig:decoder}(b). The blue square is the same decoder as a 7.4~mm$^2$ ASIC in the ASAP7 predictive 7~nm process, with its standard cells at 60\% utilization. Right: one CNN block of the routed FPGA build, about $18\times6$~mm, colored by function.}
    \label[edfigure]{fig:silicon_scale}
\end{figure*}

We map the $d=19$ decoder to the ASAP7 predictive 7~nm process~\cite{clark2016asap7} with OpenROAD~\cite{ajayi2019openroad}, at the typical corner and 0.7~V.
One CNN block routes at 0.75~ns (1.33~GHz), so at 400 cycles per round the decoder takes $0.30~\mu$s per round; at one round per microsecond the block draws about 0.55~W, and it occupies 0.42~mm$^2$ with its standard cells at 60\% utilization.
To a first approximation, the area and power of the CNN are those of this block times $L$.
Therefore, this design uses about 10~W of power (assuming one round per microsecond) and occupies 7.4~mm$^2$; contrast this with the 1{,}341~mm$^2$ of the four dies of the VP1802 (\cref{fig:silicon_scale}).

\FloatBarrier
\subsection*{Evaluation}
\phantomsection\label{meth:evaluation}

The surface and color results report the logical error rate per round $\varepsilon$, which solves $P=[1-(1-2\varepsilon)^R]/2$ for the measured failure probability $P$ of an $R$-round experiment ($R=120$, as in Ref.~\cite{alphaqubit2}).
The AQ2, AQ2-RT, Willow RT, and Chromobius values are those of Ref.~\cite{alphaqubit2}.
MWPM decodes the detector error model of each circuit with PyMatching~\cite{higgott2022pymatching,higgott2025sparse}, and union-find is a software implementation of the Helios decoder~\cite{liyanage2024fpga}.
The density of a model (\cref{fig:color_multi}(d)) is the ratio of the weights of its convolution and its two Monarch maps, to the weights of a model of the same width with depthwise convolutions and the sparsest possible Monarch maps.

The BB results report the logical error rate per logical qubit per round $\varepsilon$, which solves $q=[1-(1-2\varepsilon)^R]/2$ for the failure probability per logical qubit, $q=1-(1-P)^{1/12}$, where $P$ is the probability that any of the 12 $Z$ logicals is incorrectly decoded.
\Cref{fig:other_codes}(b) counts hardware in LUT-equivalents, LUTs plus 265 per DSP slice, since a 12-term INT8 dot product that takes four DSP58 slices takes about 1{,}060 LUTs when built from logic instead.
The Rapids points use their routed builds; Relay-BP-1 takes the 4.2 million LUTs~\cite{maurer2025relay} and no DSP slices.
XYZ-Relay-BP-5 is floating-point Relay-BP~\cite{muller2024relax} with the Relay-5 parameters of Ref.~\cite{maurer2025relay}, decoding $X$ and $Z$ jointly over the whole experiment.
X/Z-Relay-BP-1 uses the int4 arithmetic of the FPGA decoder of Ref.~\cite{maurer2025relay} in the real-time configuration: the $X$ and $Z$ checks decoded separately in windows of 12 rounds with a commit-size of 8, stopping at the first converged leg.
In this configuration it takes a median of 6 iterations per window at $p=0.1\%$, but 157 and 383 at the 99th and 99.9th percentiles, or more than $0.47$ and $1.15~\mu$s per round at 24~ns per iteration.

\FloatBarrier
\subsection*{Latency limits}
\phantomsection\label{sec:scaling}

\begin{figure*}[tbp]
    \centering
    \includegraphics[width=\linewidth]{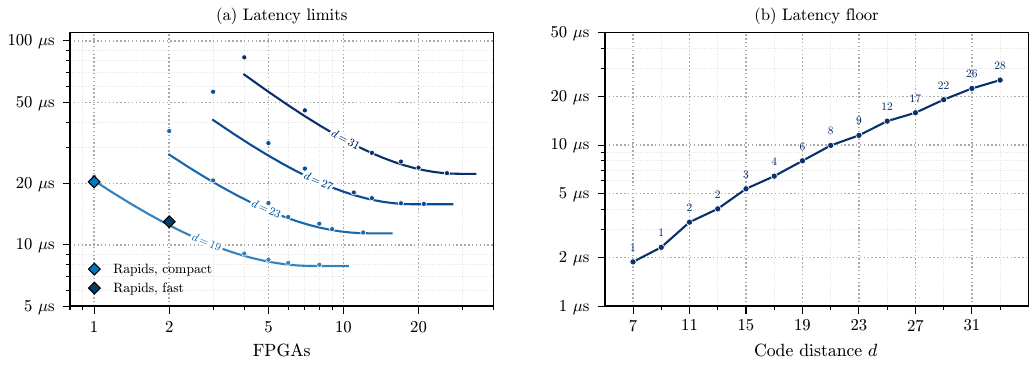}
    \caption{\textbf{Latency limits.} (a)~Latency of the surface-code decoder with a given number of FPGAs, from the model of Methods: dots, whole numbers of FPGAs; lines, fractional numbers; diamonds, the routed $d=19$ builds. Each curve flattens at the floor beyond which chip-to-chip communication dominates the latency. (b)~The floor of each curve against code distance, labelled with the fewest FPGAs that come within 5\% of it.}
    \label[edfigure]{fig:limits}
\end{figure*}

The latency limits of \cref{fig:limits} come from a model of the surface decoder, calibrated on the routed $d=19$ builds.
We assume the CNN uses $L=d-2$ blocks of width $C\propto d\log d$, with $C=64$ at $d=19$, rounded to a multiple of 16: the depth sets how far information travels, so it grows as $d$, and the width sets how much information each site holds.
At low $p$, a logical failure comes from a cluster of about $d$ defects. To pair them as matching would, a site must hold its path cost to each of these defects, a number up to about $d$ that takes about $\log d$ bits, so the information per site, and with it the width, grows as $d\log d$.
$N$ FPGAs split the CNN by depth and run $s$ sites per cycle, as many as their DSP slices allow: a block at one site per cycle uses $10C$ DSP58 slices, and we assume that 80\% of the 14{,}352 DSP slices of a VP1802 are usable before routing congestion sets in, so $s=0.8\times14{,}352\,N/(10CL)$.
A round then takes $\tau=(d+1)^2/s$ cycles, and the decision latency is
\begin{equation}
    T = \frac{L\tau + L\left(\delta_{\mathrm{block}}+\tau/(d+1)\right) + \delta_{\mathrm{out}}}{f_{\mathrm{clk}}} + (N-1)\,t_{\mathrm{link}},
    \label{eq:latency_model}
\end{equation}
where the three terms in cycles are the buffer, the CNN pipeline, and the pool, GRU and readout.
Each block adds a pipeline delay of $\delta_{\mathrm{block}}=64$ cycles and waits $\tau/(d+1)$ cycles for one row of the grid, the pool, GRU and readout take $\delta_{\mathrm{out}}=404$ cycles, $f_{\mathrm{clk}}=400$~MHz, and $t_{\mathrm{link}}=0.29~\mu$s is the delay of each chip-to-chip crossing.

\FloatBarrier
\subsection*{Use of AI tools}
We used Claude to assist in software development, including writing fast GPU kernels for training, and implementing, optimizing and debugging the FPGA builds.

\subsection*{Data availability}
The FPGA designs and the generator that produces them, the per-build records, the routed checkpoints, the floorplans, and the power and ASIC estimates behind every hardware result are available on Zenodo at \href{https://doi.org/10.5281/zenodo.24360244}{doi:10.5281/zenodo.24360244}.

\end{document}